\documentclass[12pt,a4paper,twoside]{article}
\RequirePackage{latexsym,amsmath,amssymb}
\RequirePackage[dvipsnames,usenames]{color}
\usepackage{a4wide}
\usepackage{amssymb}
\usepackage{bm}
\usepackage{dcolumn}
\usepackage{amsfonts}
\usepackage{graphicx,epsfig}
\usepackage{psfrag}
\usepackage{multicol}
\usepackage{tikz}
\usetikzlibrary{arrows,shapes}
\usetikzlibrary{matrix,arrows}
\newcommand{\be}{\begin{eqnarray}}
\newcommand{\ee}{\end{eqnarray}}
\newcommand{\bra}[1]{\mbox{$\langle\, #1 \mid$}}

\newcommand{\ket}[1]{\mbox{$\mid #1\,\rangle$}}

\newcommand{\pro}[2]{\mbox{$\langle\, #1 \mid #2\,\rangle$}}
\newcommand{\expec}[1]{\mbox{$\langle\, #1\,\rangle$}}

\renewcommand{\d}{\mbox{${\rm d}$}} 
\newcommand{\lp}{\ell_{\rm p}}
\newcommand{\mpl}{m_{\rm p}}

\newcommand{\rh}{r_{\rm H}}

\newcommand{\Rh}{R_{\rm H}}

\newcommand{\psis}{{\psi}_{\rm S}}

\newcommand{\psih}{{\psi}_{\rm H}}

\title{\bf Global and Local Horizon Quantum Mechanics}
\author{Roberto~Casadio$^{ab}$\thanks{E-mail: casadio@bo.infn.it}
$\ ,$
Andrea~Giugno$^{c}$\thanks{E-mail: A.Giugno@physik.uni-muenchen.de}
$\ $
and
Andrea~Giusti$^{ab}$\thanks{E-mail: andrea.giusti@bo.infn.it}
\\
\\
$^a${\em Dipartimento di Fisica e Astronomia, Universit\`a di Bologna}
\\
{\em via Irnerio~46, I-40126 Bologna, Italy}
\\
\\
$^b${\em I.N.F.N., Sezione di Bologna, IS - FLAG}
\\
{\em via B.~Pichat~6/2, I-40127 Bologna, Italy}
\\
\\
$^c${\em Arnold Sommerfeld Center, Ludwig-Maximilians-Universit\"at}
\\
{\em Theresienstra{\ss}e 37, 80333 M\"unchen, Germany}
}
\begin{document}
\maketitle
\begin{abstract}
Horizons are classical causal structures that arise in systems
with sharply defined energy and corresponding gravitational radius.
A global gravitational radius operator can be introduced for a static and
spherically symmetric quantum mechanical matter state by lifting
the classical ``Hamiltonian'' constraint that relates the gravitational radius
to the ADM mass, thus giving rise to a ``horizon wave-function''.
This minisuperspace-like formalism is shown here to be able to consistently
describe also the local gravitational radius related to the Misner-Sharp
mass function of the quantum source, provided its energy spectrum
is determined by spatially localised modes.
\end{abstract}
\section{Introduction}
\label{intro}
\setcounter{equation}{0}
A black hole can be viewed as a gravitationally bound state confining all possible
signals within its horizon. 
According to General Relativity, such extreme causal structures occur in very compact
gravitating systems, and understanding black holes from the perspective of particle
physics is akin to solving QCD in the strongly coupled regime.
Like the strongly coupled regime of QCD is usually investigated by means of effective
descriptions, one could analogously conceive effective quantum descriptions for
the observables of relevance in the physics of black holes, 
and processes that lead to their formation~\cite{os},
rather than insisting in (or in preparation for) the search of the ultime
quantum theory of gravity. 
\par
The semiclassical picture, in which matter fields are quantised on classical black hole manifolds,
has led to the discovery of the Hawking radiation~\cite{hawking}, and paradoxes indicating
a possibly fundamental incompatibility between the quantum theory (of fields) and General Relativity. 
Both of these pillars of physics emerged from the rethinking of the interplay between
a physical system and the observer.
It might therefore be useful to investigate first and foremost which variables
could best capture the crucial (and hopefully observable) physics of black holes,
despite of what we have come to regard as otherwise fundamental.
Examples of this approach are already found among the attempts at quantising canonically the
Einstein-Hilbert action~\cite{dirac,Bergmann,DeWitt,rovelli}.
DeWitt himself already realised the extreme complication of this programme
in his  seminal 1967 paper~\cite{DeWitt}, and immediately reverted to a simplified formulation
based on preserving isotropy and homogeneity of the universe at the quantum level.
These symmetries reduce the superspace of all possible metrics to the
Friedman-Robertson-Walker minisuperspace for the cosmic scale factor. 
\par
Black holes, and the collapse of compact objects leading to their formation~\cite{os},
cannot be realistically modelled as homogeneous systems, and their quantum
description will necessarily be more involved from the onset.
However, one can proceed similarly and consider a reduced superspace by imposing
isotropy or axial symmetry, or selecting directly a family of metrics, like, for instance,
in Refs.~\cite{kuchar,hajicek,davidson}.
Alternatively, one could study the general properties of causal structures that typically
appear in such space-times, and then quantise their intrinsic (gravitational) degrees of
freedom~\cite{ashtekar}.
In any case, it is important to remark that these ways of quantisation concern
some (reduced) degrees of freedoms, mostly in a manner independent of the
(conventional) quantum state of the matter source.
\par
In the following, we shall further develop an effective quantum description
of static horizons originally called ``horizon wave-function'' (HWF)~\cite{fuzzyh,gupf}
and later on ``horizon quantum mechanics'' (HQM) in Ref.~\cite{hqm}~\footnote{For
an attempt at including time evolution, see Refs.~\cite{hqm,Tevo}.}.
The peculiarity of this approach is that it aims at describing quantum mechanically the
existence of trapping surfaces (which reduce to horizons in the static case) from
the quantum state of the source that produces the black hole.
The HQM can therefore be viewed as complementary to approaches which
consider the horizon and black holes as purely gravitational (or metric) objects, in that
the gravitational degrees of freedom are taken ``on-shell'' with respect to a suitable
constraint with the matter state.
The HQM can then be naturally applied to systems which do not turn out to be black holes,
that is, to matter systems with a negligible probability of being black holes, or to objects
``on the verge'' of being black holes~\cite{cx}.
\par
In more details, our construction is based on the classical key concept of the gravitational radius
of a static and spherically symmetric self-gravitating source, for which this quantity
determines the existence of horizons.
We recall that we can always write a spherically symmetric metric
$g_{\mu\nu}$ as
\be
\d s^2
=
g_{ij}\,\d x^i\,\d x^j
+
r^2(x^i)\left(\d\theta^2+\sin^2\theta\,\d\phi^2\right)
\ ,
\label{metric}
\ee
where and $x^i=(x^1,x^2)$ are coordinates on surfaces of constant
$\theta$ and $\phi$.
A horizon then exists where the expansion of null geodesics vanishes,
$g^{ij}\,\nabla_i r\,\nabla_j r =0$,
$\nabla_i r$ being perpendicular to surfaces of constant area
$\mathcal{A}=4\,\pi\,r^2$.
If we set $x^1=t$ and $x^2=r$~\footnote{Let us remark this is the frame in which the 
Tolman-Oppenheimer-Volkoff equation is usually derived~\cite{tov}.},
and denote the static matter density as $\rho=\rho(r)$, Einstein equations tell us that
$g^{rr}=1-{\rh(r)}/{r}$,
where~\footnote{We shall use units with $c=1$,
and the Newton constant $G=\lp/\mpl$, where $\lp$ and $\mpl$
are the Planck length and mass, respectively, and $\hbar=\lp\,\mpl$.} 
\be
\rh(r)
=
2\,\lp\,\frac{m(r)}{\mpl}
\label{hoop}
\ee
is the gravitational radius determined by the Misner-Sharp mass function
\be
m(r)
=
4\,\pi
\int_0^r \rho(\bar r)\,\bar r^2\,\d \bar r
\ .
\label{M}
\ee
A horizon then exists where $g^{rr}=0$, or where the gravitational radius
satisfies 
\be
\rh(r)= r
\ ,
\label{Ehor}
\ee
for $r>0$.
In the vacuum outside the region where the source is located,
the Misner-Sharp mass approaches the Arnowitt-Deser-Misner (ADM) mass
of the source,
\be
\lim_{r\to\infty} m(r)
=
M
\ ,
\label{adm}
\ee
and the gravitational radius likewise becomes the Schwarzschild radius 
\be
\Rh
=
2\,\lp\,\frac{M}{\mpl}
\ .
\label{Hc}
\ee
\par
If the source is described by quantum physics, the quantities that define 
the Misner-Sharp mass $m$ (and ADM mass $M$) should become quantum
variables and one expects the gravitational radius will undergo the same
fate. 
The HQM was precisely proposed~\cite{fuzzyh} in order to describe the
``fuzzy'' Schwarzschild (or gravitational) radius of a localised (but likewise
fuzzy) quantum source.
It is important to emphasise once more that the HQM differs from most previous attempts
in which the gravitational degrees of freedom of the horizon, or of the black hole metric,
are quantised independently of the state of the source. 
In our case, the gravitational radius is instead quantised along with the matter
source that produces it, somewhat more in line with the highly non-linear general
relativistic description of the gravitational interaction in the strong regime.
\par
Clearly, the HQM becomes particularly relevant for sources of the Planck
size~\cite{cx,otherD},
for which quantum effects may not be neglected.
The Heisenberg principle of quantum mechanics introduces an uncertainty in
the particle's spatial localisation of the order of the Compton-de~Broglie length,
$\lambda_M \simeq \lp\,{\mpl}/{M}$, and $\Rh$ only makes sense if
\be
\Rh\gtrsim \lambda_M
\qquad
\Leftrightarrow
\qquad
M\gtrsim\mpl
\ .
\label{clM}
\ee
In fact, this is the argument that grants the Planck mass (and Planck length)
a remarkable role in the search for a quantum theory of
gravity~\cite{hossenfelder}.
It is comforting that the HQM predicts a particle is very likely a black hole only if 
\eqref{clM} holds~\cite{gupf,acmo,cc}.
Remarkably, it also predicts that a truly macroscopic black hole cannot be produced by
a very localised source~\cite{hqm,gupf}, but could be associated with an extended
source~\cite{dvali,BEC_BH,qhbh}.
\par
In the next Section, we shall first review the general foundations of the HQM,
and then show that the same physical states that describe the global
horizon also describe local horizons, provided the spectral decomposition
involves spatially localised energy eigenmodes.
The consequences of the discrete spectrum for a GUP and
Hawking radiation will also be briefly considered.
We shall finally comment on our results in Section~\ref{conc}.
\section{Horizon Quantum Mechanics}
\label{HQM}
\setcounter{equation}{0}
The HQM emerges from relating the matter source to its gravitational radius
at the quantum level~\cite{fuzzyh,hqm}, and allows us to put on more quantitative
grounds the condition~\eqref{clM} that should distinguish black holes from
regular particles.
\par
Before reviewing and extending this formalism, let us clarify the underlying
viewpoint by noting that one could describe formally the state of a quantum
system for which there exist two relevant sets of variables, say $X$ and $Y$, as
\be
\ket{\Psi}
=
\sum_{\alpha,\beta}
c_{\alpha\beta}\,
\ket{X_\alpha,Y_\beta}
\ .
\ee
Without loss of generality, we can group terms in the above superposition
as
\be
\ket{\Psi}
=
\sum_{\alpha,\beta}
\left(
A_{\alpha\beta}
\ket{X_\alpha,Y_\beta}
+
B_{\alpha\beta}
\ket{X_\alpha}\ket{Y_\beta;X_\alpha}
+
D_{\alpha\beta}
\ket{Y_\beta}\ket{X_\alpha;Y_\beta}
+
C_{\alpha\beta}
\ket{X_\alpha}\ket{Y_\beta}
\right)
\ ,
\label{sum}
\ee
where
\be
\hat X\ket{X_\alpha}
=
X_\alpha\ket{X_\alpha}
\quad
{\rm and}
\quad
\hat Y\ket{Y_\beta}
=
Y_\beta\ket{Y_\beta}
\ .
\label{eXY}
\ee
In the sum in Eq.~\eqref{sum}, the first term has no specific features;
the second (third) term is of the kind that admits the Born-Oppenheimer
approximation with $X$ (respectively $Y$) representing slow degrees
of freedom compared to $Y$ (respectively $X$);
finally, the fourth term contains the contribution from the direct product of
the two separate Hilbert spaces of the eigenstates~\eqref{eXY}.
We can now view $X$ as ``matter'' degrees of freedom, such as
the usual standard model fields, and $Y$ as ``gravitational'' degrees
of freedom.
Upon further assuming the state $\ket{\Psi}$ only contains
$\sum_\beta \ket{Y_\beta}\sim\ket{Y_s}$ which reproduces a (semi-)classical
configuration, the third term in Eq.~\eqref{sum} would reduce to the usual
approach of QFT on a given (curved) background~\cite{BO}, 
$\ket{\Psi}\sim \sum_\alpha \ket{Y_s}\,\ket{X_\alpha;Y_s}$,
in which only the matter fields retain their full quantum properties~\footnote{Conversely,
but perhaps of less interest, the second term would be useful in order to describe states
in which matter can be approximated classically but gravity remains
fully quantum.}.
\par
One could also do without the (semi)classical approximation.
We shall in fact see below that the states of relevance for the HQM are
of the fourth kind in the sum~\eqref{sum}, provided we suitably reduce the
matter degrees of freedom to $X=H$ (the ``matter energy'') and $Y=\Rh$
(the gravitational radius).
This is not very different from the usual quantum mechanical treatment
of the hydrogen atom, in which one quantises the (reduced) electron's position,
whereas terms containing $\ket{Y_s}$ replaced by an electron's energy level yield
the Lamb shift due to the QFT description of the Coulomb field.
\subsection{Global gravitational radius}
\label{gHQM}
We only consider spherically symmetric sources which are both localised in space
and at rest in the chosen reference frame.
If the specific source is not at rest, one should therefore change coordinates accordingly
before applying the following analysis.
We denote with $\alpha$ the set of (discrete or continuous) quantum numbers parametrising
the spectral decomposition of the source, so that our matter state can be written
as
\be
\ket{\psis}
=
\sum_\alpha
C_{\rm S}(E_\alpha)\,\ket{E_\alpha}
\ ,
\label{psis}
\ee
where the sum formally represents the spectral decomposition in Hamiltonian
eigenmodes,
\be
\hat H
=
\sum_\alpha
E_\alpha\ket{E_\alpha}\bra{E_\alpha}
\ ,
\label{HM}
\ee
and $H$ should be specified depending on the system we wish to consider.
We can then replace the r.h.s.~of Eq.~\eqref{adm} defining the ADM mass
with the expectation value of the Hamiltonian,
\be
M
\to
\bra{\psis}
\hat H
\ket{\psis}
&\!\!=\!\!&
\bra{\psis}
\sum_\alpha
E_\alpha\ket{E_\alpha}
\pro{E_\alpha}{\psis}
\nonumber
\\
&\!\!=\!\!&
\sum_{\alpha}
|C_{\rm S}(E_\alpha)|^2\,E_\alpha
\ ,
\ee
which follows from the orthonormality of the energy eigenstates,
\be
\pro{E_\alpha}{E_\beta}
=
\delta_{\alpha\beta}
\ ,
\label{ortho}
\ee
where $\delta$ is the Kronecker delta for a discrete energy spectrum and
the Dirac delta for a continuous spectrum~\footnote{This point is purely
technical in the global approach, but will become crucial in the local
analysis.}.
Let us then introduce the gravitational radius eigenstates
\be
\hat R_{\rm H}\,\ket{{\Rh}_\beta}
=
{\Rh}_\beta\,\ket{{\Rh}_\beta}
\ .
\ee
We can now show that a physical state for our system can be described by
linear combinations like the fourth term in Eq.~\eqref{sum},
\be
\ket{\Psi}
=
\sum_{\alpha,\beta}
C(E_\alpha,{\Rh}_\beta)\,
\ket{E_\alpha}
\ket{{\Rh}_\beta} 
\ .
\label{Erh}
\ee
In fact, the algebraic (Hamiltonian) constraint~\eqref{Hc} now reads
\be
0
=
\left(
\hat H
-
\frac{\mpl}{2\,\lp}\,\hat R_{\rm H}
\right)
\ket{\Psi}
=
\sum_{\alpha,\beta}
\left(
E_\alpha-\frac{\mpl}{2\,\lp}\,{\Rh}_\beta
\right)
C(E_\alpha,{\Rh}_\beta)\,
\ket{E_\alpha}
\ket{{\Rh}_\beta} 
\ ,
\ee
and is clearly solved by the combination in Eq.~\eqref{Erh} with
\be
C(E_\alpha,{\Rh}_\beta)
=
C(E_\alpha,{2\,\lp}\,E_\alpha/{\mpl})\,\delta_{\alpha\beta}
\ .
\label{solG}
\ee
\par
By tracing out the gravitational radius part, we should recover the
matter state, that is
\be
\ket{\psis}
=
\sum_\alpha
C\left(E_\alpha,{2\,\lp}\,E_\alpha/{\mpl}\right)
\ket{E_\alpha}
\ ,
\label{PsiS}
\ee
which implies
\be
C\left(E_\alpha,{2\,\lp}\,E_\alpha/{\mpl}\right)
=
C_{\rm S}(E_\alpha)
\ .
\ee
Now, by integrating out the matter states, we will obtain
\be
\ket{\psih}
=
\sum_{\alpha}
C_{\rm S}({\mpl\,{\Rh}_\alpha}/{2\,\lp})\,\ket{{\Rh}_\alpha}
\ ,
\ee
where ${\mpl\,{\Rh}_\alpha}/{2\,\lp}=E({\Rh}_\alpha)$.
We have thus recovered the HWF~\cite{fuzzyh}
\be
\psih({\Rh}_\alpha)
=
\pro{{\Rh}_\alpha}{\psih}
=
C_{\rm S}({\mpl\,{\Rh}_\alpha}/{2\,\lp})
\ ,
\label{psihd}
\ee
where the values ${\Rh}_\alpha$ form a discrete (continuous) spectrum if
$E_\alpha$ is discrete (continuous) in the quantum number $\alpha$. 
\par
If the spectral decomposition~\eqref{psis} is continuous, so will be the HWF,
and we can write
\be
\psih({\Rh})
=
\mathcal{N}_{\rm H}\,C_{\rm S}({\mpl\,{\Rh}}/{2\,\lp})
\ ,
\label{psiH}
\ee
with 
${\mathcal{N}^{-2}_{\rm H}}=4\,\pi\int_0^\infty \left|C_{\rm S}(\mpl\,{\Rh}/{2\,\lp})\right|^2 \Rh^2\,\d\Rh$
determined by the scalar product~\footnote{Note the integration
is formally extended from zero to infinity, although it will be naturally limited to a smaller
range if the spectral decomposition of the source is limited above and/or below.}
\be
\pro{\psih}{\phi_{\rm H}}
=
4\,\pi
\int_0^\infty
\psih^*({\Rh})\,\phi_{\rm H}({\Rh})
\,\Rh^2\,\d\Rh
\ .
\label{prod}
\ee
In this continuous case, the normalised wave-function~\eqref{psiH} yields the probability
density
\be
\mathcal{P}_{\rm H}(\Rh) = 4\,\pi\,\Rh^2\,|\psih(\Rh)|^2
\ee
that we would detect a gravitational radius of size $\Rh$ associated with the particle in the
quantum state $\ket{\psis}$.
Moreover, we can define the conditional probability density that the particle lies
inside its own gravitational radius $\Rh$ as
\be
\mathcal{P}_<(r<\Rh)
=
P_{\rm S}(r<\Rh)\,\mathcal{P}_{\rm H}(\Rh)
\ ,
\label{PrlessH}
\ee
where
$P_{\rm S}(r<\Rh)= 4\,\pi \int_0^{\Rh} |\psis(r)|^2\,r^2\,\d r$
is the usual probability that the particle is found inside a sphere of radius $r=\Rh$.
One can also view $\mathcal{P}_<(r<\Rh)$ as the probability density that the sphere $r=\Rh$
is a trapping surface.
Finally, the probability that the particle described by the state $\ket{\psis}$ is a
black hole (regardless of the horizon size), will be obtained by integrating~\eqref{PrlessH}
over all possible values of $\Rh$, namely
\be
P_{\rm BH}
=
\int_0^\infty
\mathcal{P}_<(r<\Rh)\,\d \Rh
\ ,
\ee
which
will depend on the observables and parameters of the specific matter state.
\subsection{Local gravitational radius}
\label{lHQMd}
We have seen that the global gravitational radius can be described irrespectively of whether
the spectral decomposition is discrete or continuous.
In order to show that the same physical quantum states~\eqref{Erh} with coefficients given in
Eq.~\eqref{solG} also allow for a local description of the gravitational radius, we shall instead
need localised energy eigenmodes and correspondingly discrete energy quantum numbers.
\par
Instead of the ADM mass~\eqref{adm}, we now start from Misner-Sharp mass at finite
radius, and again assume the system is static, so that $m=m(r)$. 
We first observe that the total Hamiltonian~\eqref{HM} associated with the ADM mass
can also be written as
\be
\hat H
&\!\!=\!\!&
\sum_{\alpha}
E_\alpha
\ket{E_\alpha}
\pro{E_\alpha}{E_\alpha}
\bra{E_\alpha}
\nonumber
\\
&\!\!=\!\!&
\sum_{\alpha}
E_\alpha
\sum_{r=0}^\infty
\ket{E_\alpha}
\pro{E_\alpha}{r}
\pro{r}{E_\alpha}
\bra{E_\alpha}
\nonumber
\\
&\!\!=\!\!&
4\,\pi
\sum_{\alpha}
E_\alpha
\int_0^\infty
\left|
\psi_{E_\alpha}(\bar r)
\right|^2
\bar r^2\,\d\bar r\, 
\ket{E_\alpha}
\bra{E_\alpha}
\ .
\label{Har}
\ee
which follows from the discrete orthogonality condition~\eqref{ortho} and the
(continuous) decomposition of the identity
$\sum_{r=0}^\infty \ket{r}\bra{r}
\equiv
4\,\pi \int_0^\infty \bar r^2\,\d\bar r\,\ket{r}\bra{r}
=
\hat{\mathbb{I}}$.
We can analogously introduce the radius-dependent Hamiltonian
\be
\hat H(r)
&\!\!=\!\!&
\sum_{\alpha}
E_\alpha
\sum_{r'=0}^r
\ket{E_\alpha}
\pro{E_\alpha}{r'}
\pro{r'}{E_\alpha}
\bra{E_\alpha}
\nonumber
\\
&\!\!=\!\!&
\sum_{\alpha}
E_\alpha\,
P_\alpha(r)\,
\ket{E_\alpha}
\bra{E_\alpha}
\ ,
\ee
where we defined
\be
P_\alpha(r)
=
4\,\pi
\int_0^r
\left|
\psi_{E_\alpha}(\bar r)
\right|^2
\bar r^2\,\d\bar r
\ ,
\ee
and note that $\lim_{r\to\infty} P_\alpha(r) = 1$.
This property only holds for square integrable (that is, {\em localised\/})
energy eigenmodes, a restriction which was not necessary in the global
case, since the norm of these modes never entered explicitly in that calculation.
However, the condition $\pro{E_\alpha}{E_\alpha}=1$ is now necessary in order to
obtain~\eqref{Har} above, which means the local construction requires the existence
of localised (bound) Hamiltonian eigenmodes.
\par
Assuming again the spectral decomposition~\eqref{psis}, the Misner-Sharp mass
function can now be replaced by the radius-dependent quantity
\be
\label{eq_1}
m(r)
\to
\bra{\psis}
\hat H(r)
\ket{\psis}
=
\sum_{\alpha}
|C_{\rm S}(E_\alpha)|^2\,
E_\alpha\,P_\alpha(r)
\ .
\ee
The gravitational radius~\eqref{hoop} analogously depends on $r$, and
we introduce the local gravitational radius eigenstates,
\be
\hat r_{\rm H}(r)\,\ket{{\Rh}_\beta}
=
{\rh}_\beta(r)\,\ket{{\Rh}_\beta}
\ .
\ee
where it is again the operator that carries a radial label.
If we now recall a physical state that satisfies the global (Hamiltonian) constraint can be written as
\be
\ket{\Psi}
=
\sum_{\alpha}
C_{\rm S}(E_\alpha)\,
\ket{E_\alpha} \ket{{\Rh}_\alpha} 
\ ,
\label{localPsi}
\ee
it immediately follows that it will also satisfy the local (Hamiltonian) constraint~\eqref{hoop}
for all values of $r$, that is
\be
0
&\!\!=\!\!&
\left[\hat H(r)
-
\frac{\mpl}{2\,\lp}\,\hat R_{\rm H}(r)
\right]
\ket{\Psi}
\nonumber
\\
&\!\!=\!\!&
\sum_{\alpha}
\left[
E_\alpha(r)-\frac{\mpl}{2\,\lp}\,{\rh}_\alpha(r)
\right]
C_{\rm S}(E_\alpha)\,
\ket{E_\alpha}
\ket{{\Rh}_\alpha}
\ ,
\ee
provided the local eigenvalues
\be
{\rh}_\alpha(r)
=
P_\alpha(r)\,
{\Rh}_\alpha
\ .
\label{solr}
\ee
Since the spectral decomposition must now be discrete, so are the above eigenvalues, and one has
\be
\bra{\psih}\hat r_{\rm H}(r)\ket{\psih}
&\!\!=\!\!&
\sum_{\alpha}
|C_{\rm S}(E_\alpha)|^2
P_\alpha(r)\,
{\Rh}_\alpha
\nonumber
\\
&\!\!=\!\!&
\sum_{\alpha}
|\psih({\Rh}_\alpha)|^2
P_\alpha(r)\,
{\Rh}_\alpha
\ ,
\label{lrh}
\ee
where $\psih$ is the (discrete) global HWF.
Finally, the classical local condition~\eqref{Ehor} for the existence of a (static) trapping surface
at the radius $r$ can now be directly replaced by 
\be
\bra{\psih}\hat r_{\rm H}(r)\ket{\psih}
=
r
\ ,
\ee
which therefore defines quantum local horizons.
\par
It is worth discussing further what would go wrong if the spectral decomposition~\eqref{psis}
did not contain only localised energy eigenmodes but also, say spatially homogenous modes,
like the plane waves.
The function $P_\alpha=P_\alpha(r)$ would increase monotonically with $r$ without bounds,
and any attempt at regularising it, for example by enclosing the system in a ``box''
of size $R$, would lead to expressions like $P_\alpha\sim r/R$
that explicitly depend on the (arbitrary) cut-off $R$, a clear sign of inconsistency.
This is of course implicitly seen already from Eq.~\eqref{Har} which again only
makes sense if $\pro{E_\alpha}{E_\alpha}=1$, as we remarked above.
\subsection{GUP and Hawking radiation}
In Ref.~\cite{gupf}, a GUP was obtained by combining (linearly) the uncertainty in the source size
$\expec{\Delta \hat r^2}$ encoded in the matter state~\eqref{PsiS} with the uncertainty in the horizon size
$\expec{\Delta \hat R_{\rm H}^2}$ given by a (continuous) global HWF~\eqref{psiH}.
In particular, for Gaussian matter states, 
\be
\psis(r)
\simeq
e^{-\frac{r^2}{\ell^2}}
\ ,
\label{gauss}
\ee
where $\ell\simeq \lp\,\mpl/m$ is the Compton width of the source of mass $m$,
the GUP was shown to take the form
\be
\Delta r
\simeq
\lp\,\frac{\mpl}{\Delta p}
+\gamma\,\lp\,\frac{\Delta p}{\mpl}
\ ,
\label{gupG}
\ee
with $\gamma$ an arbitrary coefficient, 
\be
\expec{\Delta \hat r^2}
=
4\,\pi\int_0^\infty
|\psis(r)|^2\,r^4\,\d r
-
\left(4\,\pi\int_0^\infty
|\psis(r)|^2\,r^3\,\d r
\right)^2
\simeq
\ell^2
\ ,
\ee
and
\be
\expec{\Delta \hat R_{\rm H}^2}
=
4\,\pi\int_0^\infty
|\psih(\Rh)|^2\,\Rh^4\,\d \Rh
-
\left(4\,\pi\int_0^\infty
|\psih(\Rh)|^2\,\Rh^3\,\d \Rh
\right)^2
\simeq
\frac{\lp^4}{\ell^2}
\ .
\ee
Finally, the global uncertainty in radial momentum is given by
\be
\Delta p^2
=
4\,\pi\int_0^\infty
|\psis(p)|^2\,p^4\,\d p
-
\left(4\,\pi\int_0^\infty
|\psis(p)|^2\,p^3\,\d p
\right)^2
\simeq
\mpl^2\,\frac{\lp^2}{\ell^2}
\ ,
\ee
where
\be
\psis(p)
\simeq
e^{-\frac{p^2}{m^2}}
\ .
\ee
\par
As we have just shown, were one to employ the local description of Section~\ref{lHQMd}, the spectrum of the source 
should be discrete, and consequently so would be the HWF. 
In particular, we can now define a local uncertainty in the source size~\footnote{More technically,
one can view $4\,\pi\,\expec{\Delta \hat r^2(r)}^2$ as the uncertainty in the area of a sphere of coordinate
radius $r$.}
\be
\expec{\Delta \hat r^2(r)}
=
4\,\pi\int_0^r
|\psis(\bar r)|^2\,\bar r^4\,\d \bar r
-
\left(4\,\pi\int_0^r
|\psis(\bar r)|^2\,\bar r^3\,\d \bar r
\right)^2
\ .
\label{ldr}
\ee
Likewise, and replacing integrals with sums over the spectral index $\alpha$, we can also
define local uncertainties for the gravitational radius
\be
\expec{\Delta \hat r_{\rm H}^2(r)}
=
\sum_\alpha
|C_{\rm S}(E_\alpha)|^2\,{\Rh}_\alpha^2\,P_\alpha(r)
-
\left(
\sum_\alpha
|C_{\rm S}(E_\alpha)|^2\,{\Rh}_\alpha\,P_\alpha(r)
\right)^2
\ ,
\label{ldrh}
\ee
and for the radial momentum 
\be
\Delta p^2(r)
=
\sum_\alpha
|C_{\rm S}(E_\alpha)|^2\,p^2_\alpha\,P_\alpha(r)
-
\left(\sum_\alpha
|C_{\rm S}(E_\alpha)|^2\,p_\alpha\,P_\alpha(r)
\right)^2
\ ,
\ee
where $p_\alpha=p(E_\alpha)$.
\par
It is interesting to note that, for spectral eigenmodes, the above sums would reduce to single terms
and one finds
\be
\frac{\expec{\Delta \hat r_{\rm H}^2(r)}}{{\Rh}_\alpha^2}
=
P_\alpha(r)
\left[1-P_\alpha(r)\right]
=
\frac{\Delta p^2(r)}{p^2_\alpha}
\ .
\ee
By again combining linearly the size uncertainty~\eqref{ldr} with the uncertainty in the gravitational radius~\eqref{ldrh},
one obtains a local GUP of the form~\eqref{gupG} at each (finite) value of $r$.
Since $P_\alpha(r\to\infty)=1$, one also finds
$\expec{\Delta \hat r_{\rm H}^2(r\to\infty)}=\expec{\Delta \hat R_{\rm H}^2}=0=\Delta p^2(r\to\infty)$,
in agreement with the fact that the global GUP reduces to the standard Heisenberg uncertainty relation for spectral 
eigenmodes.
Of course, Gaussian states~\eqref{gauss} may now not be in the discrete Hilbert space~\footnote{Of course,
they might still be obtained as the limit of suitable series.},
or they could instead be energy eigenstates, depending on the details of the system at hand. 
In any case, it is hard to conceive that macroscopic black holes are simple spectral eigenmodes,
and some form of global GUP should therefore apply.
\par
A more drastic consequence follows for the Hawking radiation, as one expects 
only quanta corresponding to transitions between states of the discrete spectrum
be allowed.
Moreover, this argument would further support the idea that macroscopic black holes cannot be
spectral eigenmodes, as those would not support any Hawking emission.
It is in fact tempting to draw a connection with corpuscular models of black holes~\cite{dvali},
as they appear to be bound states in the sense shown in Ref.~\cite{baryons}, and do not
suffer of the paradoxes related to the standard description of the Hawking radiation.
\section{Conclusions}
\label{conc}
\setcounter{equation}{0}
We have analysed the quantum constraint that relates the gravitational radius
of a spherically symmetric source to its spectral decomposition, and shown that the
same quantum state can be employed in order to describe both the global radius
associated with the ADM mass and the local radius associated with the Misner-Sharp
mass function.
\par
A crucial difference that emerges between the local and global gravitational radius at
the quantum level is that the former requires the spectral decomposition is done in terms
of localised energy eigenmodes, whereas the global radius can be defined in any case.
From the physical point of view, one can argue the global gravitational radius
is an asymptotic property of a self-gravitating system and should therefore be rather
insensitive to the details of its internal structure, whereas the local gravitational radius
should be determined by the precise internal structure of the source. 
It therefore appears consistent that the local gravitational radius can be defined only
provided the inner structure of the source is properly characterised as well.
Finally, the fact the spectral decomposition must be discrete does not constitute a real
limitation in most practical situations, since any realistic astrophysical sources, like stars,
should have very finely-spaced energy levels.
\par
So far we have only considered the formal extension of the HQM to the local gravitational
radius, and some general considerations regarding the GUP and Hawking radiation, 
but it will next be important to apply this approach to specific models of
self-gravitating objects, and also to compare our findings with specific proposals of
quantum black holes~\cite{davidson,cx,dvali,qhbh,piero,euro}. 
%
%
%\section*{Acknowledgments}
%
%
%
%\appendix
%
%
%

%
\end{document}